\documentclass[twocolumn,trackchanges]{aastex62}

\makeatletter
\newcommand*{\centerfloat}{%
  \parindent \z@
  \leftskip \z@ \@plus 1fil \@minus \textwidth
  \rightskip\leftskip
  \parfillskip \z@skip}
\makeatother

\usepackage[utf8]{inputenc}
\usepackage[english]{babel}
\usepackage{enumitem}

\usepackage{listings}
\usepackage{xcolor}
 
\definecolor{codegreen}{rgb}{0,0.6,0}
\definecolor{codegray}{rgb}{0.5,0.5,0.5}
\definecolor{codepurple}{rgb}{0.58,0,0.82}
\definecolor{backcolour}{rgb}{0.95,0.95,0.92}

\lstdefinestyle{mystyle}{
    backgroundcolor=\color{backcolour},   
    commentstyle=\color{codegreen},
    keywordstyle=\color{magenta},
    numberstyle=\tiny\color{codegray},
    stringstyle=\color{codepurple},
    basicstyle=\ttfamily\footnotesize,
    breakatwhitespace=false,         
    breaklines=true,                 
    captionpos=b,                    
    keepspaces=true,                 
    numbers=left,                    
    numbersep=5pt,                  
    showspaces=false,                
    showstringspaces=false,
    showtabs=false,                  
    tabsize=2
}
 
\graphicspath{{./}{figures/}}

\received{XX YY ZZ}
\revised{XX YY ZZ}
\accepted{\today}

\submitjournal{PASP}

\shorttitle{Automated Extended Aperture Photometry}

\shortauthors{B\'odi et al.}

\begin{document}

\title{Automated Extended Aperture Photometry of K2 variable stars}

\author[0000-0002-8585-4544]{Attila B\'odi}
\affiliation{Konkoly Observatory, Research Centre for Astronomy and Earth Sciences, E\"otv\"os Lor\'and Research Network (ELKH), Konkoly Thege Mikl\'os \'ut 15-17, H-1121 Budapest, Hungary}
\affiliation{MTA CSFK Lend\"ulet Near-Field Cosmology Research Group, 1121, Budapest, Konkoly Thege Mikl\'os \'ut 15-17, Hungary}
\affiliation{ELTE E\"otv\"os Lor\'and University, Institute of Physics, 1117, P\'azm\'any P\'eter s\'et\'any 1/A, Budapest, Hungary}

\author[0000-0002-5781-1926]{P\'al Szab\'o}
\affiliation{Konkoly Observatory, Research Centre for Astronomy and Earth Sciences, E\"otv\"os Lor\'and Research Network (ELKH), Konkoly Thege Mikl\'os \'ut 15-17, H-1121 Budapest, Hungary}
\affiliation{MTA CSFK Lend\"ulet Near-Field Cosmology Research Group, 1121, Budapest, Konkoly Thege Mikl\'os \'ut 15-17, Hungary}

\author[0000-0002-5481-3352]{Emese Plachy}
\affiliation{Konkoly Observatory, Research Centre for Astronomy and Earth Sciences, E\"otv\"os Lor\'and Research Network (ELKH), Konkoly Thege Mikl\'os \'ut 15-17, H-1121 Budapest, Hungary}
\affiliation{MTA CSFK Lend\"ulet Near-Field Cosmology Research Group, 1121, Budapest, Konkoly Thege Mikl\'os \'ut 15-17, Hungary}
\affiliation{ELTE E\"otv\"os Lor\'and University, Institute of Physics, 1117, P\'azm\'any P\'eter s\'et\'any 1/A, Budapest, Hungary}

\author[0000-0002-8159-1599]{L\'aszl\'o Moln\'ar}
\affiliation{Konkoly Observatory, Research Centre for Astronomy and Earth Sciences, E\"otv\"os Lor\'and Research Network (ELKH), Konkoly Thege Mikl\'os \'ut 15-17, H-1121 Budapest, Hungary}
\affiliation{MTA CSFK Lend\"ulet Near-Field Cosmology Research Group, 1121, Budapest, Konkoly Thege Mikl\'os \'ut 15-17, Hungary}
\affiliation{ELTE E\"otv\"os Lor\'and University, Institute of Physics, 1117, P\'azm\'any P\'eter s\'et\'any 1/A, Budapest, Hungary}

\author[0000-0002-3258-1909]{R\'obert Szab\'o}
\affiliation{Konkoly Observatory, Research Centre for Astronomy and Earth Sciences, E\"otv\"os Lor\'and Research Network (ELKH), Konkoly Thege Mikl\'os \'ut 15-17, H-1121 Budapest, Hungary}
\affiliation{MTA CSFK Lend\"ulet Near-Field Cosmology Research Group, 1121, Budapest, Konkoly Thege Mikl\'os \'ut 15-17, Hungary}
\affiliation{ELTE E\"otv\"os Lor\'and University, Institute of Physics, 1117, P\'azm\'any P\'eter s\'et\'any 1/A, Budapest, Hungary}

\begin{abstract}

Proper photometric data are challenging to obtain in the K2 mission of the \textit{Kepler} space telescope due to strong systematics caused by the two-wheel-mode operation. It is especially true for variable stars wherein physical phenomena occur on timescales similar to the instrumental signals.
We originally developed a method with the aim to extend the photometric aperture to be able to compensate the motion of the telescope which we named Extended Aperture Photometry (EAP). Here we present the outline of the automatized version of the EAP method, an open-source pipeline called \texttt{autoEAP}. We compare the light curve products to other photometric solutions for examples chosen from  high-amplitude variable stars. Besides the photometry, we developed a new detrending method, which is based on phase dispersion minimization and is able to eliminate long-term instrumental signals for periodic variable stars. 

\end{abstract}


\section{Introduction} \label{sec:intro}

High-precision photometry collected by the \textit{Kepler} space telescope revolutionized several fields of astronomy \citep{borucki2016}, such as our understanding of exoplanets \citep{Borucki10,Batalha13,Thompson2018} and variable stars \citep{Gilliland10,Derekas-2011,Prsa11,Molnar16,Yu20}, and even discovering highly unusual phenomena \citep{Boyajian16,Rappaport-2019}. The revolution did not stop even after the subsequent failure of two reaction wheels, eventually leading to the remarkably successful K2 mission \citep{k2mission}, that also enabled the observation of moving objects in the ecliptic plane \citep{Szabo15}. However, new problems arise from the loss of precise pointing ability of the spacecraft among all axes, and the data became strongly affected by the rolling and frequent corrective attitude control maneuvers of the spacecraft. As the telescope is rolling around its optical axis, the adverse effects are nearly negligible in the center of the field-of-view, but towards the edges relative movements of stars may reach 1--2 pixels. Besides the Simple Aperture Photometry (SAP) and Pre-search Data Conditioning SAP (PDCSAP) outputs that were created for the original mission \citep{sap}, many other photometric pipelines were developed \citep{k2sff,k2p2,k2sc, everest,k2varcat} approaching the problem from two angles: the optimization of apertures and post-processing methods to correct the systematics in the data. 

These pipelines typically provide solutions for certain needs, focusing on, for example, exoplanet detection, stellar variability, or even separation of close stars, and thus give us useful data for a large fraction of the observed stars. Nevertheless, K2 photometry processing presents enough challenges to remain an active field of research years after the end of the mission \citep{bulgephot,eap,deblending}. The \texttt{lightkurve} \citep{lightkurve} package has also been developed where a custom aperture can be set manually, and therefore is extremely useful for stars where the available pipelines fail.

Such objects are for example the RR Lyraes stars, where the problem arises from two reasons: the timescales of light variability are very similar to the timescales of systematics, and the pulsation cycles are similar in shape to the sawtooth-like photometric variability induced by the attitude position variation. This resemblance confuses many of the correction pipelines. Thousands of RR Lyrae stars were observed during the mission: our group developed the method of Extended Aperture Photometry (EAP) in order to obtain useful photometry for them. Light curves of more than four hundred RR Lyrae stars of the early K2 campaigns have already been published by \citet{eap}, where apertures were determined individually and manually. That experience led us to develop the automated version of the pipeline, which we present in this paper.

This work is structured as follows: in Section \ref{sec:code} we describe each step of the code, and present our solution to define the apertures while avoiding contamination. In Section \ref{sec:results} we present example light curves and compare them to other pipelines and discuss the output quality. Our conclusions are summarized in Section \ref{sec:concl}.

\section{The pipeline workflow} \label{sec:code}

Our python-based pipeline processes the Target Pixel
Files (TPFs), defines and optimizes the pixel aperture and produces raw EAP light curves. We also implemented additional techniques that can be optionally used to further correct the raw data, the K2SC (K2 Systematics Correction) method \citep{k2sc} and a detrending algorithm of a polynomial fitting technique based on phase dispersion minimization. 
The workflow of the main part of the method, getting the raw light curve from the TPFs is illustrated in Fig. \ref{fig:workflow}. The details of each step are given below. 

\begin{figure}
    \centering
    \includegraphics[width=\columnwidth]{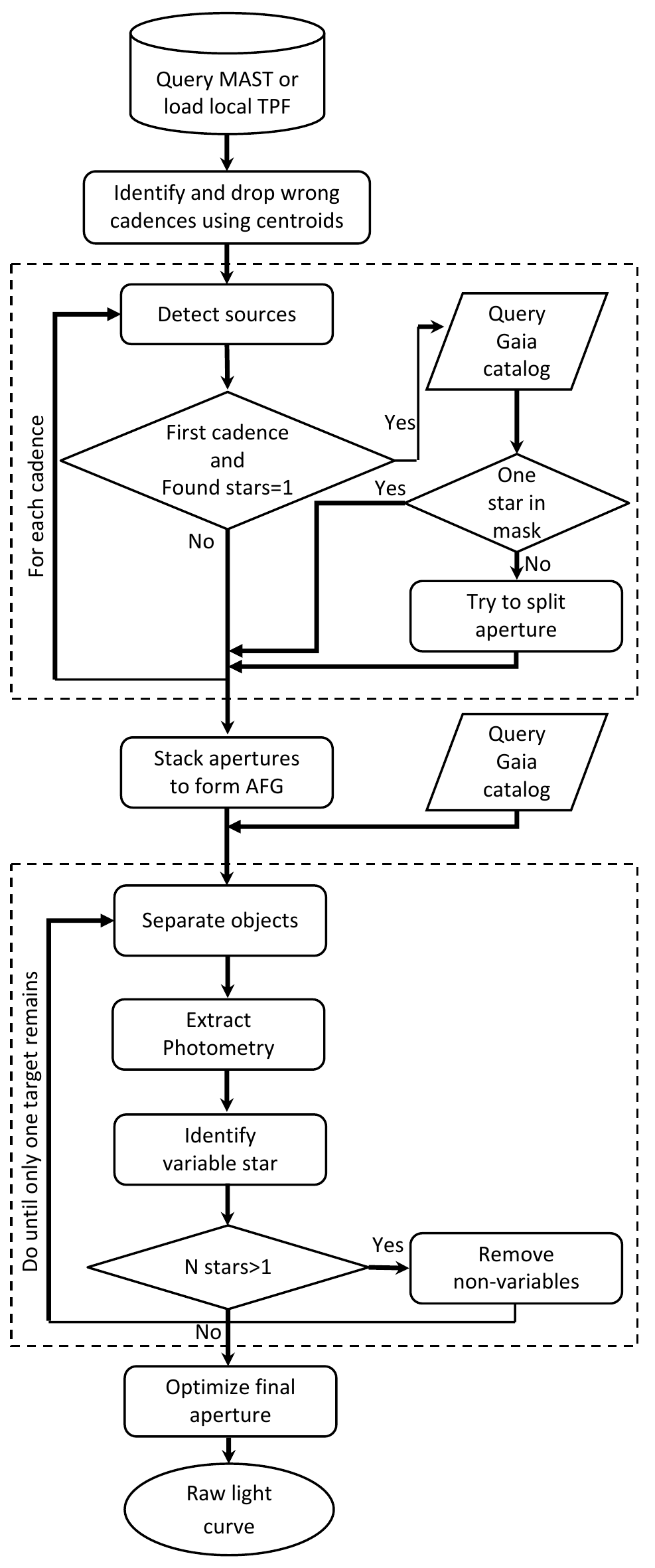}
    \caption{The workflow of \texttt{autoEAP} that shows the main steps of obtaining raw, uncorrected light curves from target pixels files. The black dashed boxes depict the two loop and iterative steps.}
    \label{fig:workflow}
\end{figure}

\subsection{Target pixel file pre-processing} \label{sec:tpf}

\begin{figure*}
    \centering
    \includegraphics[width=\textwidth]{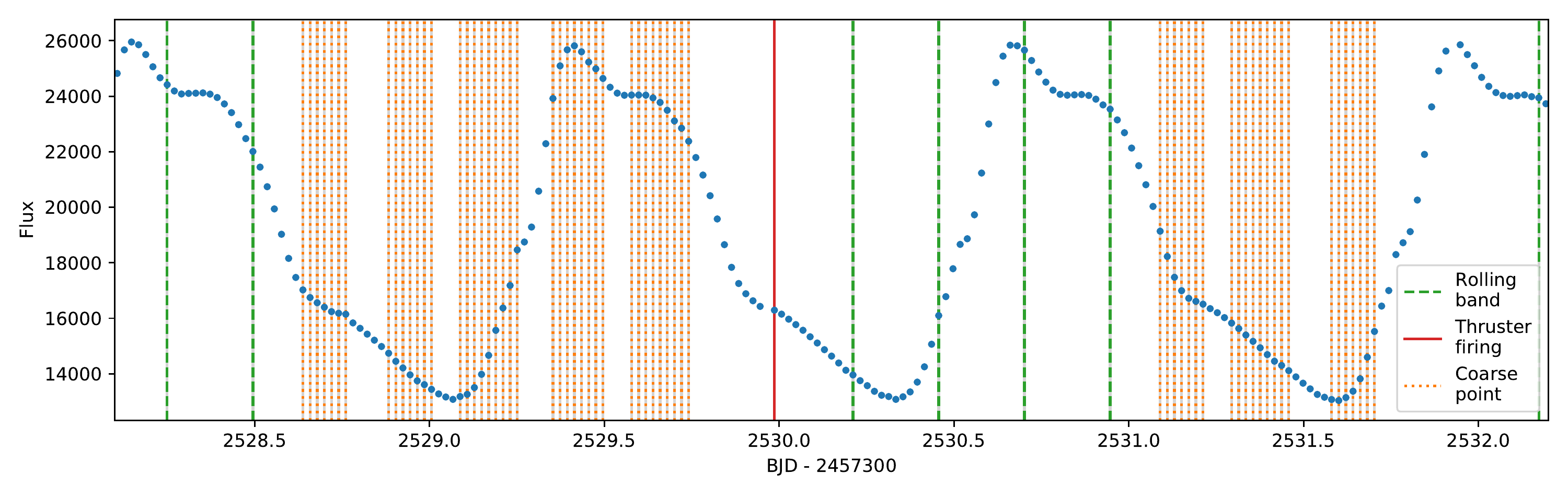}
    \caption{ An example light curve of EPIC 217235287, where we marked the cadences flagged with various quality flag bits that we found useful for photometry: flag bits 18-19 (dashed green lines), flag bit 21 (red lines), flag bit 3 (orange dotted lines).   }
    \label{fig:flags}
\end{figure*}

Due to limited storage and bandwidth allocation, K2 observations are downloaded and stored only for designated targets and corresponding pre-selected pixels, similar to the nominal \textit{Kepler} mission. The collection of active pixels around a given a star is called a target pixel file (TPF).

FITS files containing the 30-minute cadence TPFs and the corresponding metadata can be automatically downloaded from the Mikulski Archive for Space Telescopes (MAST\footnote{\url{https://archive.stsci.edu/}}) or pre-downloaded local copies can be passed to our code. The files are handled through the \texttt{TargetPixelFile} module defined in the \texttt{lightkurve} package \citep{lightkurve}, thus the built-in methods can be utilized within \texttt{autoEAP}. Each TPF consists of 3--4000 cadences, depending on the given campaign, where each cadence is a grid of pixels with raw and background-corrected flux data. The cadences are labeled with quality flags, where non-zero values refer to different kinds of data anomalies, such as problems with telescope pointing or failures that caused data loss (see the \textit{Kepler} Archive Manual for the meaning of quality bits; \citealt{KeplerArchiveManual}). From the list of quality flags, we only consider two, the one that indicates that the spacecraft was not in fine point (flag bit 16) and the one that labels cadences where no data were collected (flag bit 17). The main reason behind this decision is that several observations are flagged as `coarse point', which refers to cadences where the pointing of the telescope drifted by more than 0.5 millipixels from the nominal value \citep{KeplerDataCarHandbook}. However, in these cases the target star is usually still on the TPF and the photometry can be performed safely (see Fig.~\ref{fig:flags}). The resulting photometric issues will then be corrected by K2SC (see Sect.~\ref{sec:k2sc}). If we were to consider coarse pointing flag, which is the suggested operation by default, then several valuable light curve points would be lost.

\begin{figure}
    \centering
    \includegraphics[width=\columnwidth]{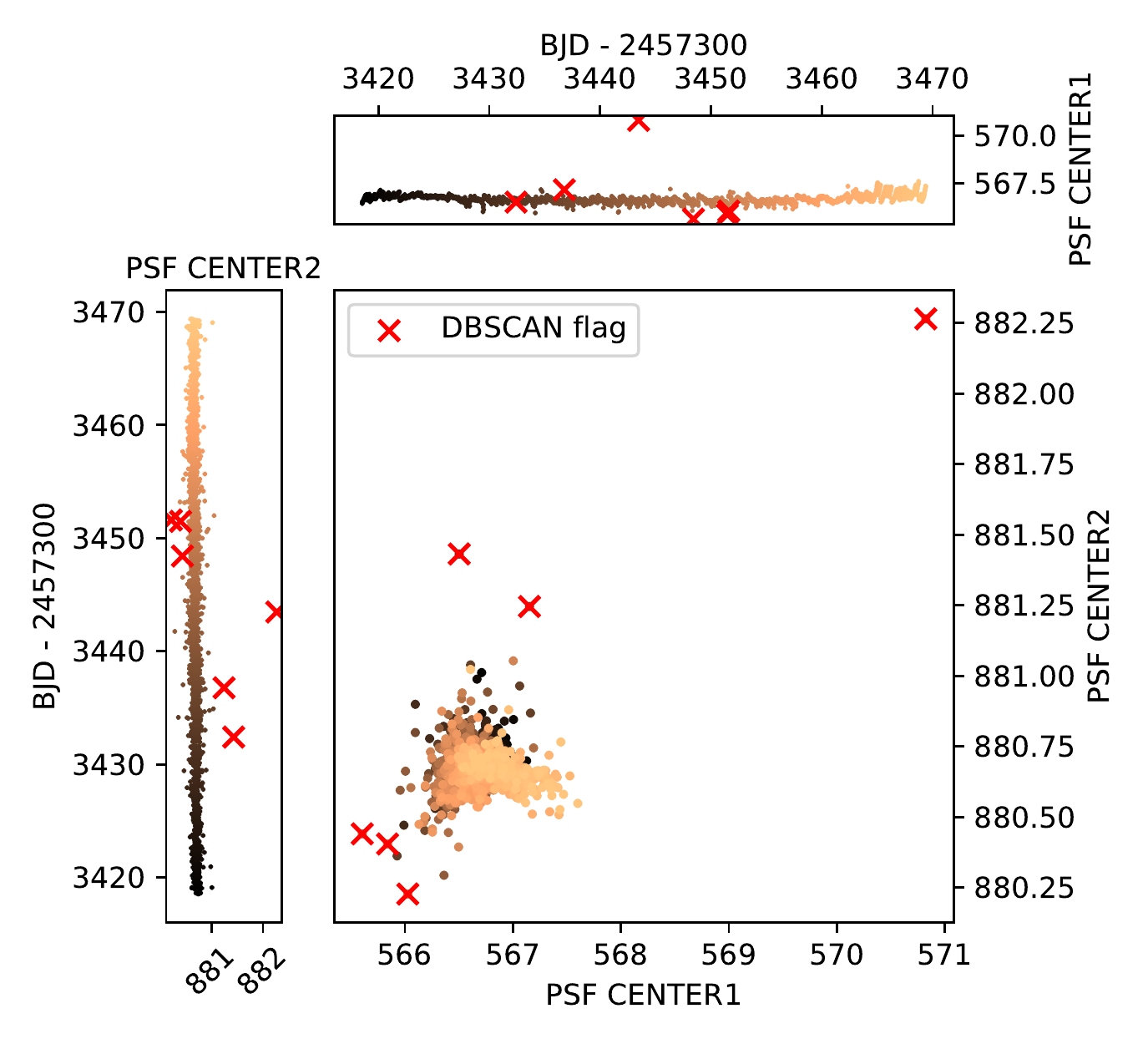}
    \caption{Photocenter movement over time for EPIC 251812081, observed in Campaign 18. \textit{Upper}: variation along the x axis. \textit{Left}: variation along the y axis. \textit{Middle}: PSF movement on the TPF. The color scale indicates the time evolution. Outliers flagged by DBSCAN are marked with red crosses.}
    \label{fig:outliers}
\end{figure}

Nonetheless, due to the excess motion of the telescope, the position of the stars cannot be approximated as constant with time. To identify outlier cadences where the central position of the stars' point spread function (PSF) was too far from its initial value, we flag points manually by computing the geometric center of each image using the \texttt{estimate\_centroids} method implemented in \texttt{lightkurve}. We determine a central cluster of centroids and separate outliers using the `Density-based spatial clustering of applications with noise' (DBSCAN) algorithm from the \texttt{scikit-learn} python package \citep{scikit-learn}. Figure~\ref{fig:outliers} shows an example for the star EPIC 251812081 that was observed in Campaign 18. It can be clearly seen that the telescope pointing varied within a range of 1-2 pixels. In most cases, very few (less than 5) cadences are classified as outliers, and in many cases, none is classified as such.

\subsubsection{Defining initial apertures} \label{sec:apdef}

\begin{figure}
    \centering
    \includegraphics[width=\columnwidth]{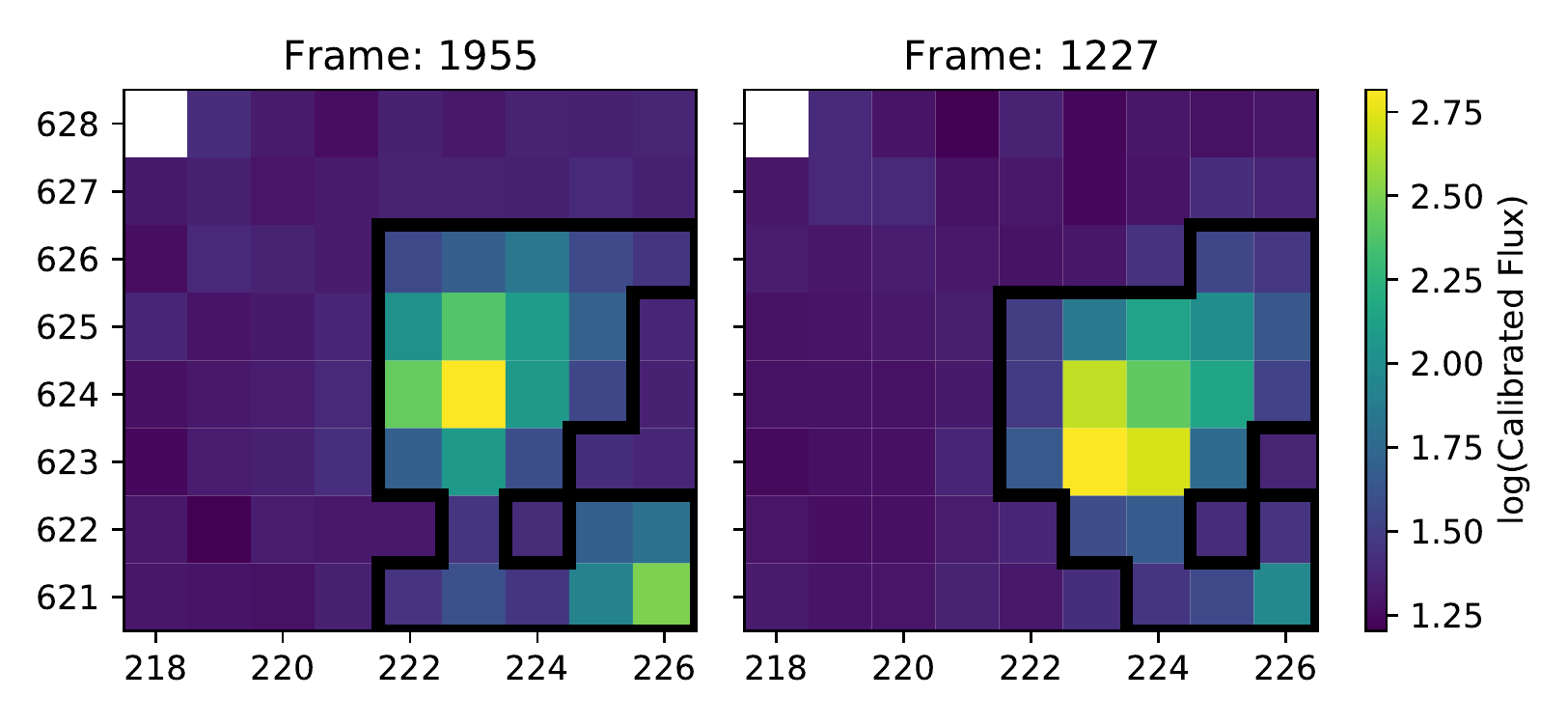}
    \caption{Two cadences with selected apertures for EPIC 246427702. The two figures show the TPF at the extremes of the telescope pointings. The target star is in the center, and the aim is to create a large aperture for its photometry with minimum contribution from the contaminating stars at the bottom of the images. Note the logarithmic flux scale.}
    \label{fig:aperture}
\end{figure}

For every non-outlier cadence, we identify sources via image segmentation using the \texttt{photutils} package from AstroPy \citep{photutils}. The identification is a two step procedure, before which the saturated pixels are masked out. Firstly, a 2D detection threshold image is produced using simple sigma-clipped statistics  with the \texttt{sigma\_clipped\_stats} algorithm of AstroPy, setting a 3-sigma threshold to estimate the background level and its RMS variation. Then the pixels that are above the background by a given threshold level are connected along their edges. This method is not able to de-blend or separate close sources on its own. Therefore if only one target is found, we use the Gaia DR2 source catalog \citep{Gaia,GaiaDR2} to check whether the initial aperture includes two or more stars that are merged together. If the latter is the case, the threshold level is iteratively altered until the sources are separated. However, this may fail in specific cases, in which the initial aperture is kept. Moreover, if the aperture contains nearby (within 3 pixels) stars of similar brightness or faint targets close to the detection limit of the \textit{Kepler} CCD, this step is skipped. Fig.~\ref{fig:aperture} shows an example cadence with the defined per-cadence apertures, where two TPFs are plotted at the extremes of the telescope pointing to illustrate the effect of the two reaction wheel controlled motion.

\begin{figure}
\centering
\includegraphics[width=\columnwidth]{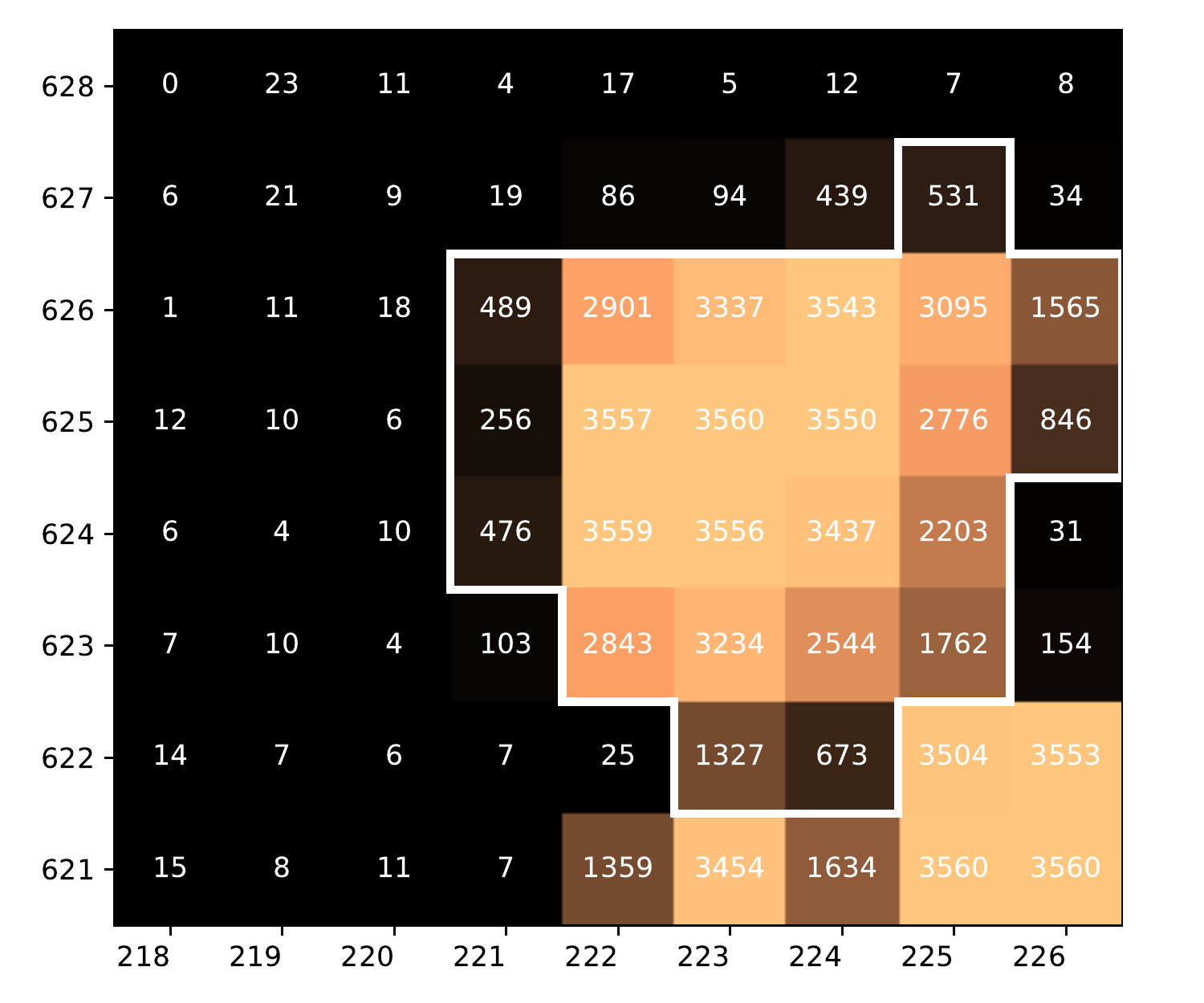}
\caption{The so-called Aperture Frequency Grid (AFG). Numbers on each pixel correspond to the number of cadences when the given pixel was part of an aperture. Structures belonging to individual stars are split in each iterative step, and AFG values of identified objects which are separate from the variable star are set to zero. The iteration stops when only one object remains. The white line depicts the edges of the final extended aperture used for photometry.}
\label{fig:afg}
\end{figure}

Having calculated apertures for all individual cadences classified as non-outliers, we count for each pixel the number of times it was selected to be part of an aperture. From these stacked apertures, we create an aperture frequency grid (AFG), which will serve as the base to form the final extended aperture, as instead of defining apertures for each cadence, our goal is to create one large aperture only that covers the target star at each time step. In the AFG, the values are between 0 and the number of cadences, which vary from campaign to campaign. An example AFG can be seen in Fig.~\ref{fig:afg}, which was formed by stacking the per-cadence apertures corresponding to Fig.~\ref{fig:aperture}. The white line depicts the final extended aperture, which was defined by separating the objects via an iterative process. The steps of this process are described in the following two subsections.

\subsubsection{Identifying targets in the AFG} \label{sec:afg}

After the AFG is defined, the next step is to identify and separate potential targets. This is done via an iterative image segmentation. The process starts with defining a threshold number. If the AFG value of a given pixel exceeds this number, then the pixel is considered to be part of an aperture, otherwise it is ignored. For the image segmentation, we use the \texttt{label} method, from \texttt{scipy}'s multidimensional image processing package \citep{scipy}. The method uses centrosymmetric matrices to separate structures in the AFG. Two apertures are different if the pixels from which they were created do not border each other on the sides. The number of apertures found this way is dependent on the threshold value. Therefore, we introduce a procedure to balance the number of targets and the selected pixel area per target, in order to not divide the TPF into too many portions, and to keep the selected area per target as large as possible to maximize the signal to noise ratio and to capture all incident fluxes.

\begin{figure}
    \centering
    \includegraphics[width=\columnwidth]{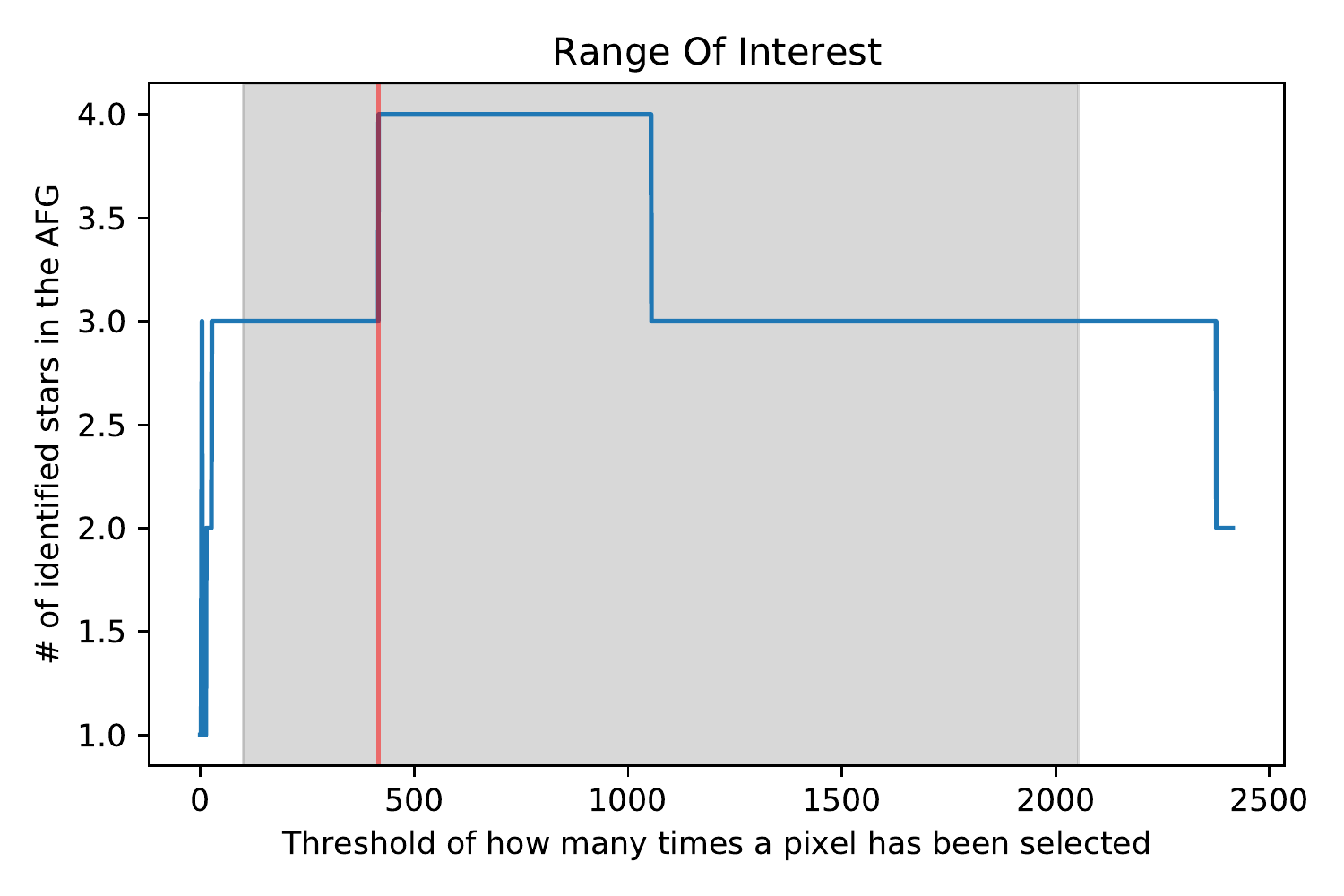}
    \includegraphics[width=\columnwidth]{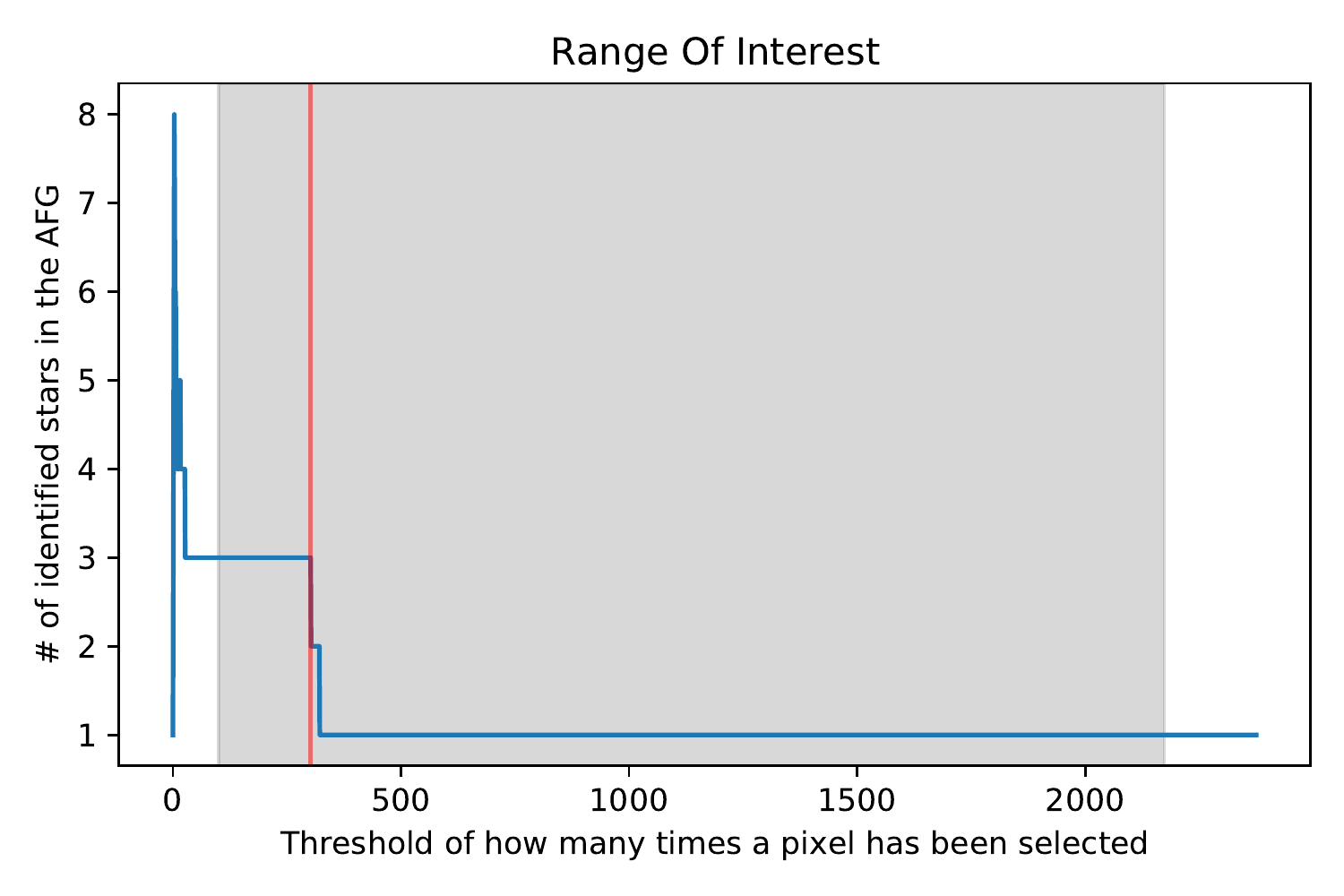}
    \includegraphics[width=\columnwidth]{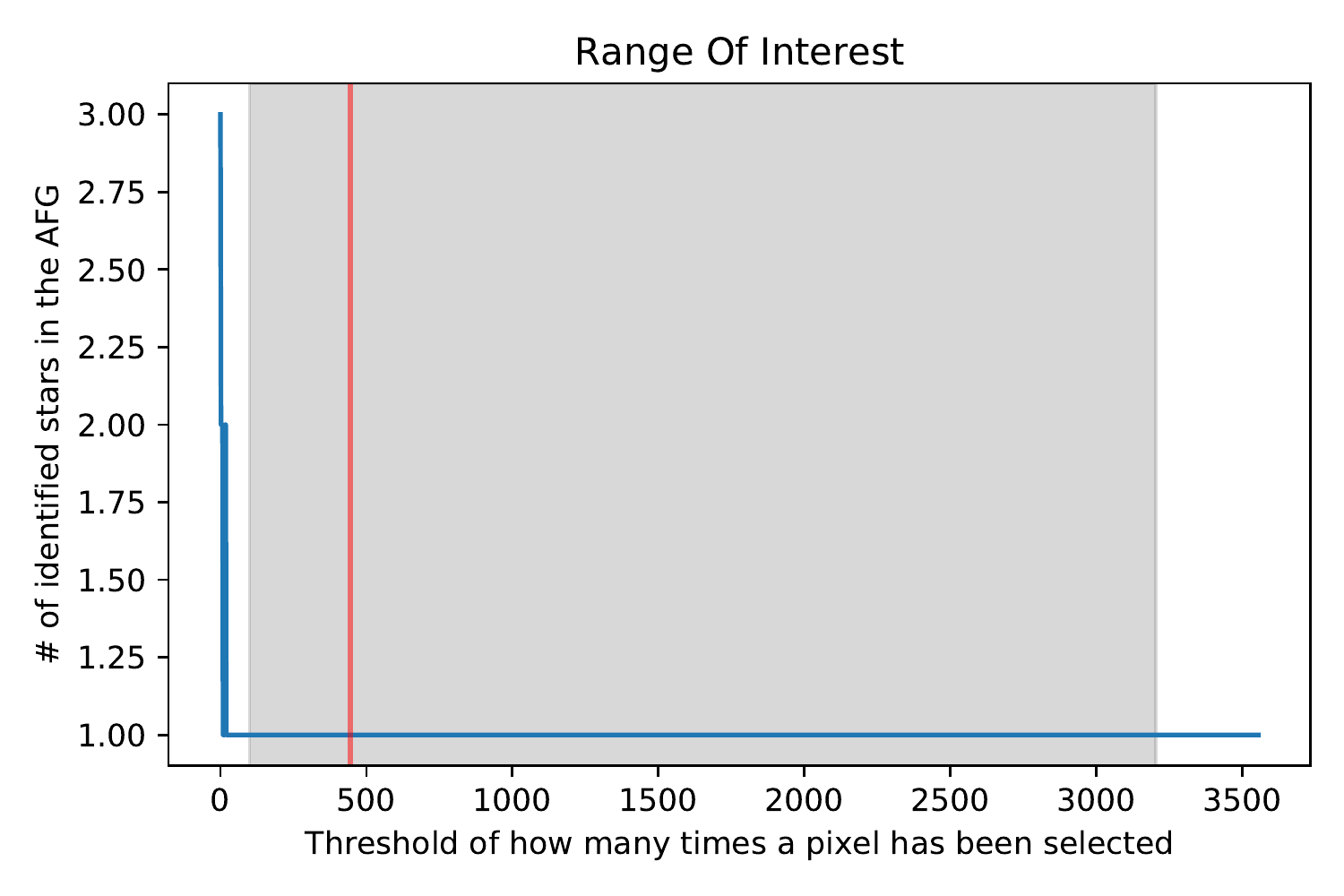}
    \caption{Number of stars identified in the AFG versus the threshold of minimal AFG value above which an AFG pixel is considered to form an aperture. The gray shaded area shows the ROI within which the final threshold is selected. \textit{Top:} If there is a break point where the number of stars increases as the threshold increases, the threshold is selected to divide the AFG into higher number of stars. Here it is four. \textit{Middle:} If the number of identified stars decreases as the threshold increases, the curve is reversed and the threshold will be from right to left at the second upward jump. \textit{Bottom:} If there is no break point, i.e., there is only one target, then the threshold is set to the number of cadences divided by the TH value to form a large aperture. See text for more details.}
    \label{fig:threshold_selection}
\end{figure}

We count the number of identified stars as a function of threshold and analyze the result. Fig.~\ref{fig:threshold_selection} shows the three possible cases. After extensive examination of how the method works with different thresholds, we set a range of interest (ROI) from which the final threshold is selected. The limits are the hundredth cadence and 85th percentile of the cadences (the latter varies from campaign to campaign). Using a wider range often results in undesired behaviour. Lower threshold may split the AFG into too many parts using the majority or even all the pixels, while a higher value would select only a small number of pixels to form apertures.

Based on the shape of the curve, the following are used to define the threshold:

\begin{enumerate}[label=(\alph*)]
\item First we look for the lowest threshold (within the ROI) that still separates stars which would not be separated if the threshold would have been decreased by 1 (see top panel of Fig.~\ref{fig:threshold_selection}). If there is a second upward jump at a higher threshold before which there is no downward jump, then the latter one is preferred. The goal of this is to achieve the largest possible aperture for a single star without another star being in the same aperture.
\item If the number of apertures within the ROI is not constant, but there is no upward jump to separate two stars, but there is a break point where the number of stars varies, we do the same as before, but first reversing the direction of the curve. This case is illustrated in the middle panel of Fig.~\ref{fig:threshold_selection}.
\item  If the number of apertures are the same for all possible thresholds, then we first extend the upper ROI limit to see if we can find another upward jump. If not, we use a pre-defined value for the threshold. This pre-defined threshold is the number of non-outlier cadences divided by a given number, called TH, which value was selected experimentally to be 8 (this case can be seen in the bottom panel of Fig.~\ref{fig:threshold_selection}).
\end{enumerate}

This method is not perfect as it assumes that all sources can be separated photometrically. Due to the strong PSF movement there are cases when two or more source are close enough to be included in one aperture. To overcome this problem, we use the Gaia DR2 source catalog again, similar to what is presented in Sect. \ref{sec:apdef} to make sure that close-by but separable stars are isolated. However, in this case the sources are forced to be separated by assigning pixels from the proposed aperture to each target based on the distance between the location of the target and the pixels weighted by the brightness of the star.

\subsubsection{Identification of variable stars}

The selected apertures are used to produce light curves using the \texttt{to\_lightcurve} method of \texttt{lightkurve}. If there are multiple apertures for one TPF, we need to find the one that belongs to the high-amplitude variable star. This is needed as we found that in some cases, especially for large FOVs, the location of the variable star is not in agreement with the EPIC coordinates. The light curves are transformed into frequency spectra using the Lomb-Scargle (LS) method from the \texttt{AstroPy} package \citep{lomb,Scargle}. The star with the highest power value is nominated as our variable star of interest. However, before this step, some confusing signals which may break this analysis  must be removed. Firstly, before the LS signal calculation, we pre-whiten with a low-frequency component to eliminate long-term trends. Secondly, as the attitude control maneuvers are also periodic their effect manifests as a strong peak in the spectrum corresponding to the $6$-hour long roll motion and subsequent reset of the telescope (with a frequency of $\approx4.07$ c/d): this signal and its harmonics are also masked out.

After identifying the variable star, the pixels in the AFG that are part of apertures of other stars are set to zero. This way we make sure that those pixels will not be considered in the final aperture (e.g., pixels with large AFG values outside of the white-line bordered area in Fig.~\ref{fig:afg}). Afterwards, the previous process from Sect. \ref{sec:afg} is repeated: a new AFG is made, the number of stars versus the threshold number is calculated and the threshold is defined, the aperture of the variable star is identified via LS analysis. This process is iterated until only one star remains. Finally, possible gaps inside the proposed aperture are revealed and filled.

\subsubsection{Fine-tuning the proposed aperture}
The original idea behind the EAP method is to conserve the flux of the
star to the highest possible extent. In \citet{eap}, in order to find the ideal target pixel masks the authors assigned the pixels individually to the stars by hand. This way they were able to make sure that the best quality light curve was achieved, but the process required considerable manual labor. To check whether the automatically selected pixels form the optimal aperture or the addition of adjacent pixels would improve the quality, we need a metric to characterize the data. We decided to use the variance given by Phase Dispersion Minimization (PDM; \citealt{PDM}), which is usually used to find the best period that minimizes the scatter of the phase-folded data compared to the overall scatter. Here we use this method to compare the scatter of different light curves extracted with different apertures by fixing the period.

\begin{figure}
    \centering
    \includegraphics[width=\columnwidth]{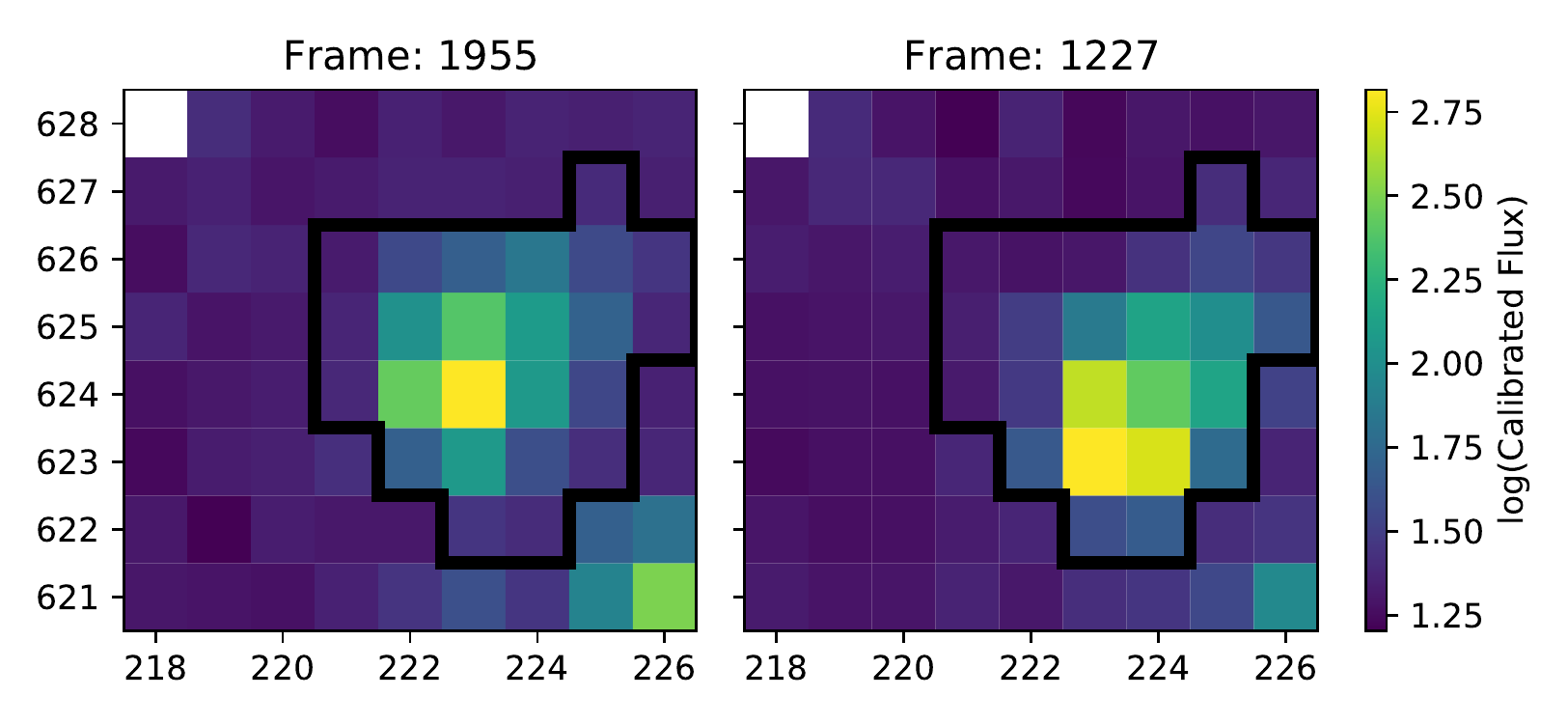}
    \caption{The final aperture that is used for the photometry of EPIC 246427702. The two frames show TPFs at the telescope pointing extremes. Per-pixel flux is shown with color, on a logarithmic scale.}
    \label{fig:finalap}
\end{figure}

To fine-tune the proposed aperture, we select and append each adjacent pixel one-by-one, perform the photometry, calculate the metrics and compare them to the initial value from the proposed aperture. To compare the PDM variance, first a Lomb-Scargle periodogram is calculated using the original light curve to estimate the period of the variable star and to phase-fold the data. The PDM variances are compared using the same period. The variance always decreases if the light curve quality improves. To avoid overextending the apertures, the new PDM values are only considered if the variance decreases by at least 0.002 compared to the initial value.
In the case of EPIC 24642770, the last aperture is shown in Fig.~\ref{fig:finalap}, where the two TPFs show the images of the star at two extreme attitude positions. As it can be seen, the contaminating stars in the bottom are left out and the target PSF is always within the aperture.

\subsubsection{Systematics corrections} \label{sec:k2sc}

The final light curve captures all or most of the flux from the target star, minimizing the systematical variations due to the variable flux loss caused by the attitude changes of the telescope. While this is often the strongest systematic present in raw K2 light curves, it is not the only one, and further instrumental effects still remain in the data, which mainly arise from three factors. First, sensitivity of adjacent pixels are different, and there is no detailed, pixel-level flat field available for the \textit{Kepler} CCD to correct for these. Sensitivity also changes within each pixel, dropping towards the pixel edges, which affects the recorded flux levels if the PSF moves about. Finally, the defined apertures are discrete, described by an integer number of pixels instead of a continuous polynomial, which would more accurately follow the shape of the PSF \citep{KeplerPRF}. We therefore perform a correction on the light curves after the photometry to remove systematics correlated with the telescope roll motion by using a modified version of the K2SC software \citep{Aigrain15,k2sc}. This is the most time consuming step in the whole procedure. K2SC uses Gaussian process regression to separate time-dependent and position-dependent signals in the light curve. The former is modelled by a quasi-periodic kernel, which is only possible if the period is known. Otherwise, a simple squared-exponential kernel is used. As the time scale of the pulsation is similar to the frequency of the attitude control maneuvers of the telescope, we modified the algorithm to reliably find the right pulsation frequency. After this we found that K2SC works effectively on stars which have a period close to 6~h, such as RR Lyraes.

The position-dependent component requires the knowledge of the telescope pointing as a function of time. The FITS file headers contain the position correction values, but in the case of cadences that are flagged as coarse points these values are missing. However, as we mention in Sect. \ref{sec:tpf}, it does not mean that those light curve points are useless. To keep as many cadences as possible, we estimate the TPF centroids using the \texttt{estimate\_centroids} method again, this time with the final aperture. After the correction is done, we calculate the median absolute deviation (MAD) and compare it to the MAD of the raw \texttt{autoEAP} light curve. If corrections based on these centroids yield worse light curves than the raw ones, i.e. if the MAD after the K2SC correction is at least 1.5 times the (outliers introduced) or less than 60\% of the raw data (intrinsic variations degraded or removed), then K2SC is rerun, but this time using the position correction values present in the TPF. Unfortunately, this means that in those cases several flagged cadences are dropped from the corrected data.

As it is described by \citet{k2sc}, the behaviour of the systematics changes qualitatively at one or two points during each campaign as the direction of the telescope roll changes. Because of this effect, the affected timestamps must be set as break-points for the position-dependent kernel. These are not pre-set for some campaigns in K2SC, but we identified these points visually and added them to the code.

\begin{figure*}
    \centering
    \includegraphics[width=\textwidth]{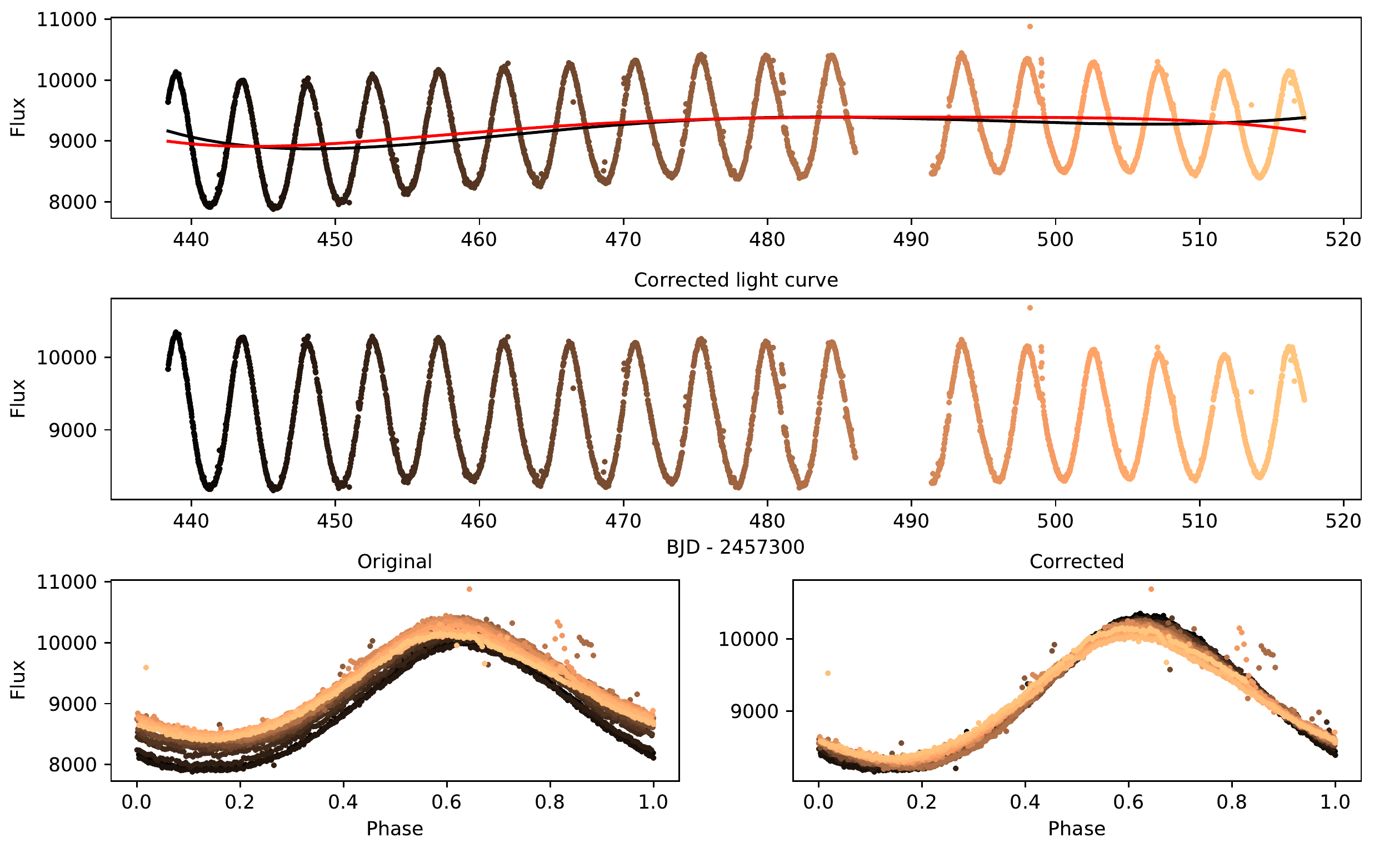}
    \caption{Phase dispersion minimization-based detrending of a spotted star, EPIC246455590, observed in Campaign 12. \textit{Top:} The K2SC corrected \texttt{autoEAP} light curve. The black line is a polynomial fitted by minimizing the deviation between the data and the curve. The red is a same-order polynomial, but after minimizing for the PDM variance. \textit{Middle:} The detrended light curve. \textit{Bottom:} The two panels show the phase-folded light curve before (left) and after (right) applying the correction. The color scale shows the time evolution, going from darker to brighter colors. Units for Barycentric Julian Dates (BJD) are in days, for fluxes they are in $e^-/s$.}
    \label{fig:PDMdetrend}
\end{figure*}

\begin{figure*}
    \centering
    \includegraphics[width=\textwidth]{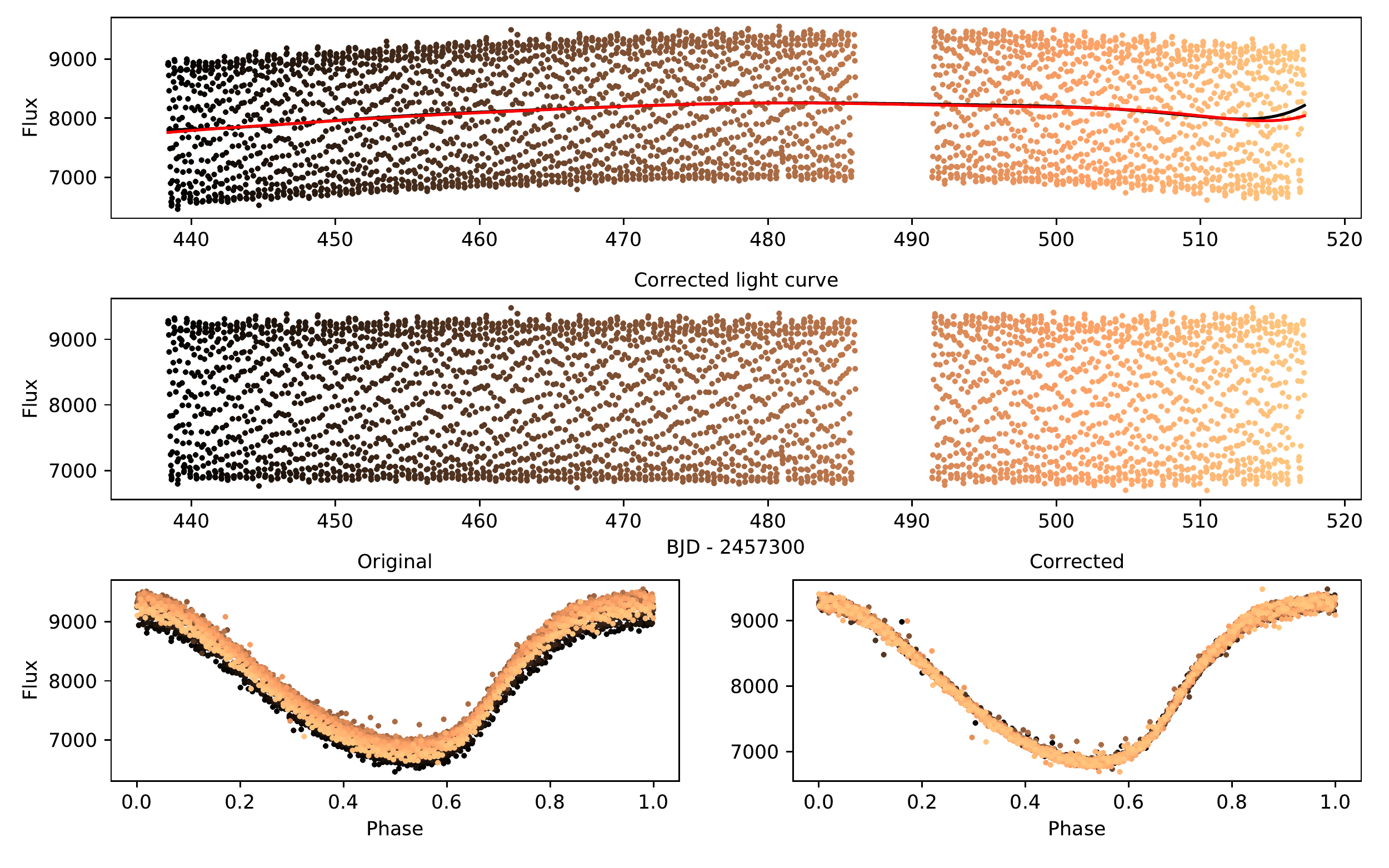}
    \caption{Phase dispersion minimization-based detrending of an RRc star, EPIC246463406, observed in Campaign 12. The panels are the same as in Fig~\ref{fig:PDMdetrend}.} 
    \label{fig:PDMdetrend2}
\end{figure*}

\subsubsection{Detrending}

The systematics correction removes instrumental signals that correlate with the movements of the telescope, but it is not well-suited to correct slow changes if they are present. When we inspected the results we found that low-frequency instrumental variations usually still remain, with shapes that differ from star to star. This effect can originate from several causes, and can potentially be intrinsic. However, due to its long-term nature it is nearly impossible to clearly connect it either to a physical reason or instrumental effects, such as contamination from nearby stars or reflections, crosstalk in the electronics, or poor background determination. One such problematic phenomenon in RR Lyrae stars is the Blazhko effect \citep{blazhko,jurcsik-2009}. In \citet{eap}, the authors checked all stars and cleaned the light curves of non-Blazhko start with strong trends by fitting and removing cubic spline curves. They left the Blazhko modulated stars untouched, as these show average brightness variation with modulation periods similar to the instrumental time scales, thus simple spline fits would have interfered with the intrinsic variations as well.

In this paper, we present a new idea to detrend the light curves of periodic variable stars, while preserving the intrinsic variations including the Blazhko effect. The technique is based on variance that is given by phase dispersion minimisation and illustrated in Figs.~\ref{fig:PDMdetrend} and \ref{fig:PDMdetrend2}, where we plotted the light curve of a spotted star and an RR Lyrae variable with longer and shorter periods, respectively, compared to the data duration. In the first step the period of the dominant variation is determined from a Lomb-Scargle periodogram. Then the light curve is smoothed with a median sliding window, where the length is proportional to the period, using the \texttt{flatten} method from the \texttt{W{\={o}}tan} package \citep{wotan}. After removing the fit from the original data, distant outliers are removed via sigma-clipping. This preparatory step is required to avoid outliers misleading the PDM calculation.

The slowly varying systematics are estimated by a low-order polynomial. To find the best-fitting curve, most algorithms minimize the difference between the fit and the data set, e.g. by calculating the $\chi^2$ statistics. The result of this approach can be seen in the upper panel of both Fig.~\ref{fig:PDMdetrend} and \ref{fig:PDMdetrend2}, where the black curve is fitted via $\chi^2$ minimization. The problem is that at the light curve edges the distribution of points is not symmetric around the mean and the fit tries to follow the tails instead of the mean. One trivial solution is to use lower-order polynomials, but in that case instrumental variations often cannot be followed properly. Instead of the $\chi^2$, our method minimizes the variance given by PDM at the previously determined period, while adjusting the free parameters of the polynomial with given order using the Powell method from \texttt{scipy}'s \texttt{optimize} \citep{scipy}. The result of this fit is shown as a red curve in the top panels of the figures. The original and corrected phase-folded light curves are shown in the bottom panels. The color scale follows time, the lighter the color the later the date. It is clearly visible that the vertical dispersion is smaller, while the horizontal variations are not affected after the correction. This PDM-based detrending technique is also able to preserve intrinsic, Blazhko-like modulations as well as phase modulations due to spot evolution or binarity.

The only parameter that has to be set is the order of the polynomial. On the one hand, if the number of covered cycles is very low, higher-order polynomials can remove intrinsic variations. On the other hand, for shorter-period stars where several cycles are observed, higher order fits can safely remove any instrumental variation. In our code, the order of the fitted polynomial is determined based on the number of the covered cycles within a range from zero (in which case the trend is a constant) to nine.

\subsubsection{Standalone detrender}

The task of detrending light curves of periodic variable stars is not restricted to K2 data. Other space-based photometric observations are also exposed to instrumental variations. Moreover, the reprocessing of raw or K2SC corrected \texttt{autoEAP} light curves becomes a lot more efficient if the photometry itself does not need to be repeated each time. Because of these considerations, we built \texttt{autoEAP} in a modular way, and the PDM-based polynomial fitting and detrending can be used as a standalone module, to be applied to any light curve data. The usage of the standalone detrender module is described in the GitHub repository\footnote{\url{https://github.com/konkolyseismolab/autoeap}} of this pipeline.

\begin{figure}
    \centering
    \includegraphics[width=\columnwidth]{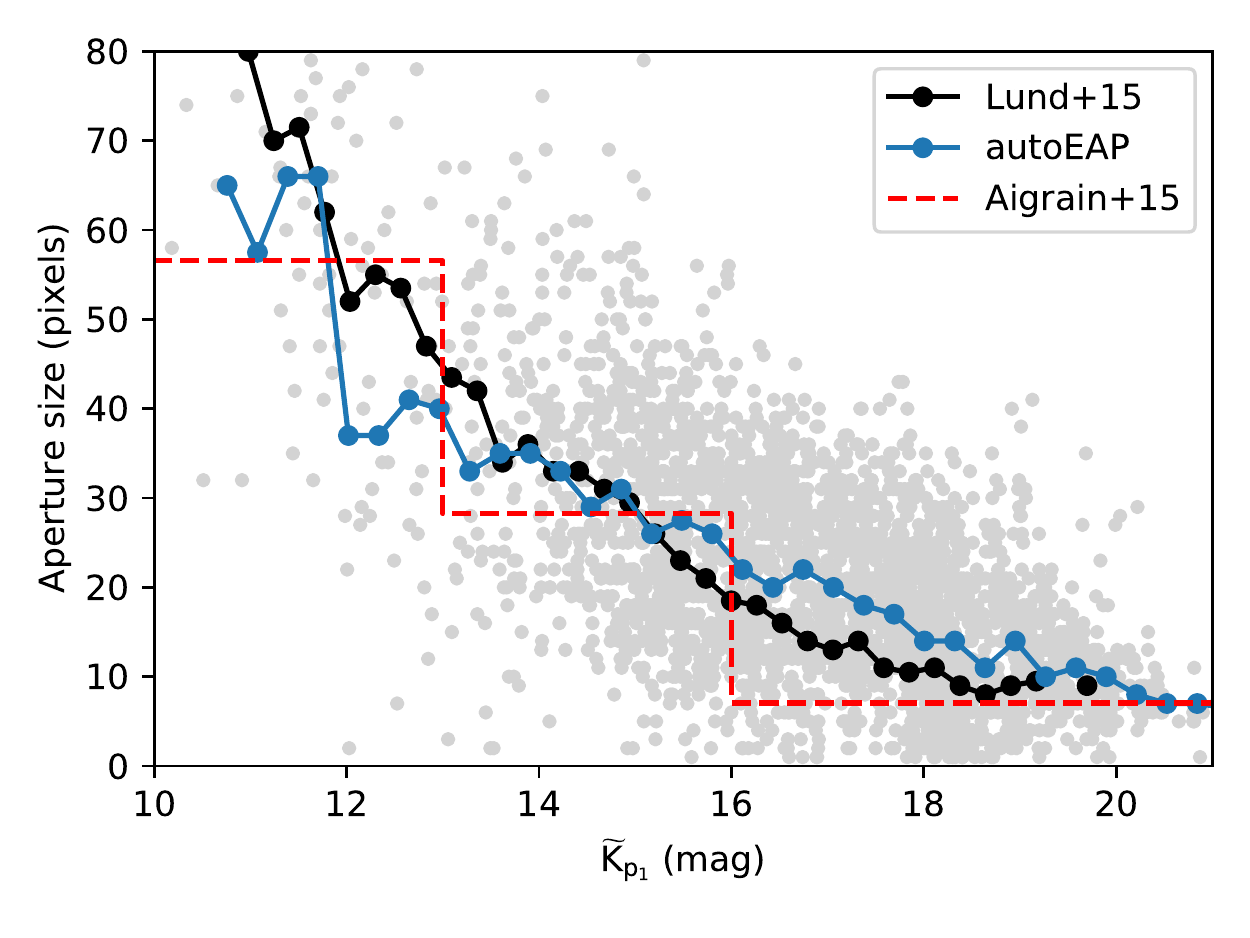}
    \caption{ Comparison of aperture sizes used for photometry as a function of brightness in magnitude. Gray points are RR Lyrae stars processed with \texttt{autoEAP}. The blue and black connected points are the median aperture sizes for \texttt{autoEAP} and \citet{k2p2}, respectively. The red dashed line marks the aperture sizes used by \citet{Aigrain15}.  }
    \label{fig:apsizes}
\end{figure}

\section{Comparison to other pipelines}
\label{sec:results}

In this section we compare our results to those of other pipelines. First we demonstrate the difference between the \texttt{autoEAP} apertures and the apertures determined by \citet{Aigrain15} and \citet{k2p2}. To compare how the aperture sizes vary with magnitude, we calculate the median brightness of the stars following \citet{k2p2}, where the $\tilde{K}_{p_1}$ magnitude is defined as
\begin{equation}
    \tilde{K}_{p_1} = 25.3 - 2.5 \log_{10}(S),
\end{equation}
where $S$ is the median flux of the light curve, measured in e$^-$/s. 

The method of \citet{k2p2} constructs the apertures in a way similar to ours. The main difference is that they stack the TPFs first, before calculating the apertures, while in the \texttt{autoEAP} method apertures are defined for the individual cadences and then the individual apertures are stacked, ensuring that each cadence contributes with equal weight, and thus high-amplitude phases do not dominate the final aperture. This leads to a clear difference between the average aperture sizes for the faint stars ($>15$ mag), where \texttt{autoEAP} apertures are systematically larger, as shown in Fig. \ref{fig:apsizes}. The routine of \citet{Aigrain15} uses circular apertures with different radii over a small number of brightness intervals.

\begin{figure*}
    \begin{center}

        \centerfloat
        \includegraphics[width=\textwidth]{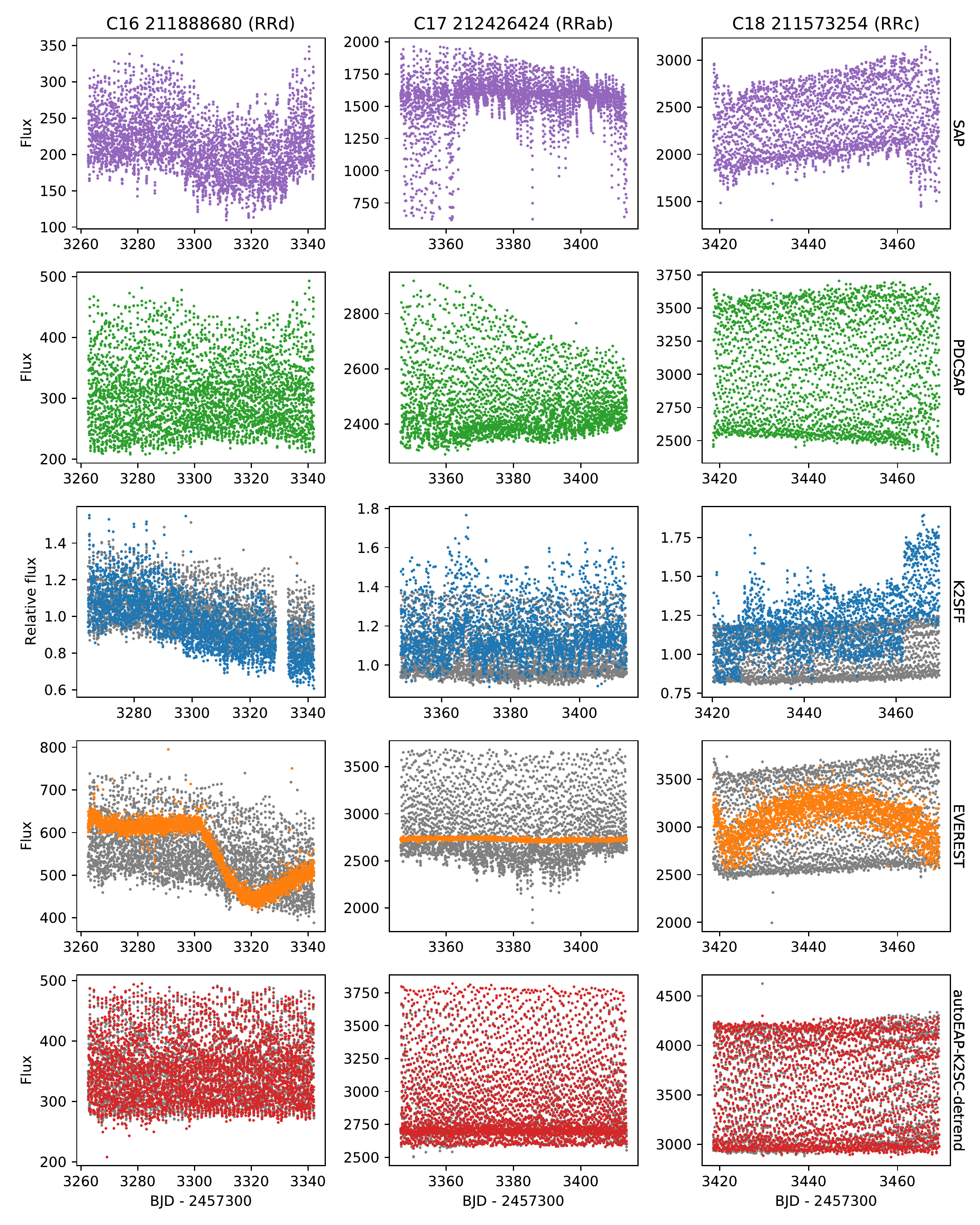}
        \caption{Comparison of RRab variable light curves produced by different pipelines for three targets at different brightness levels. Each column correspond to one star. Different rows show the result of different pipelines as indicated by the right-hand side labels. Gray points show the raw data sets of corresponding method. The \texttt{autoEAP} is in the bottom row. We applied sigma clip statistics to remove significant outliers and to improve the visibility of the light curves. Fluxes are in e$^{-}$/s units.}
        \label{fig:exampleLCs}
    \end{center}
    \end{figure*}

\begin{figure*}
    \begin{center}

        \centerfloat
        \includegraphics[width=\textwidth]{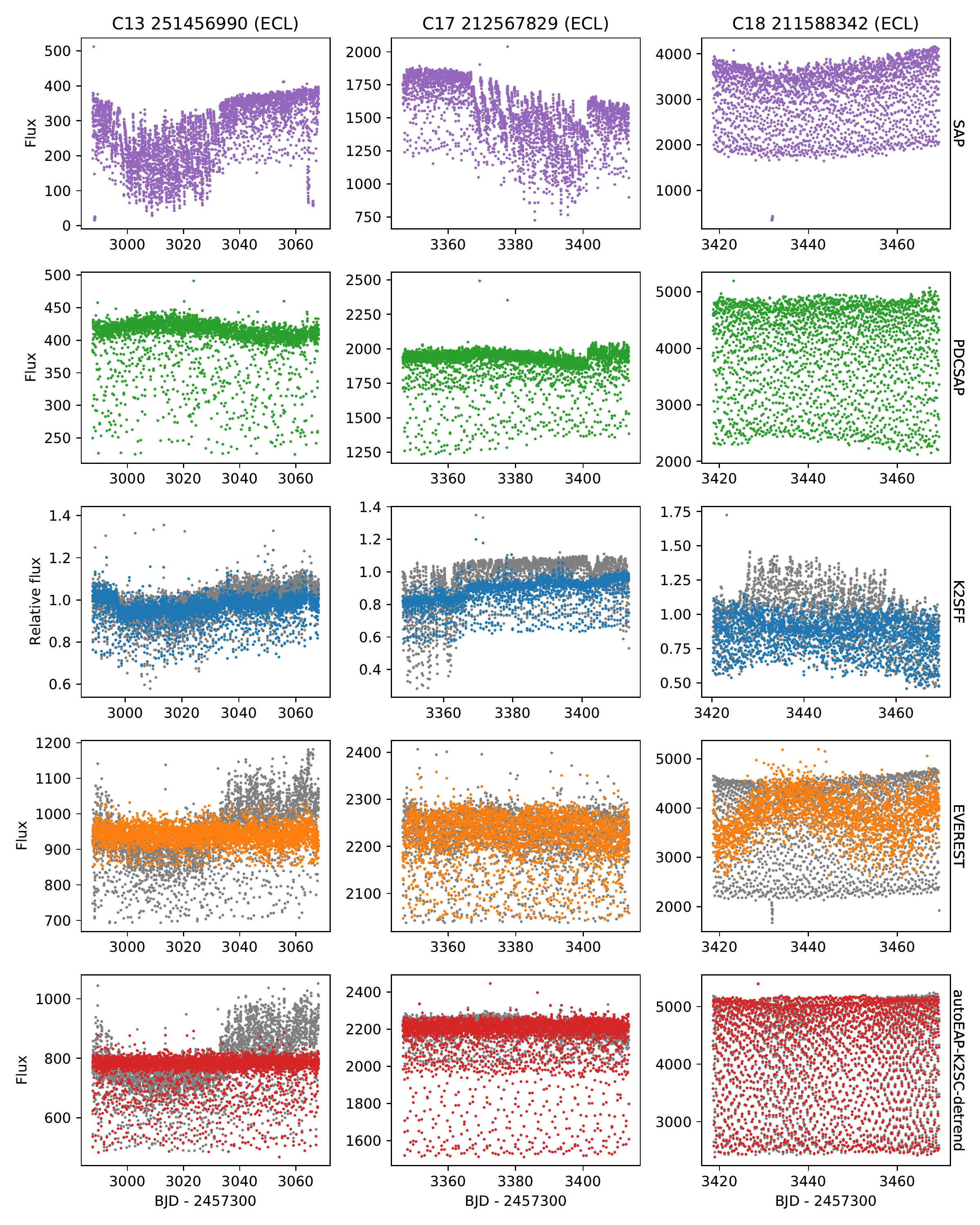}
        \caption{Comparison of eclipsing binary star light curves produced by different pipelines for three targets at different brightness levels. The structure of the figure and its labels are identical to Fig. \ref{fig:exampleLCs}.}
        \label{fig:exampleLCs2}
    \end{center}
    \end{figure*}

We also present the comparison between light curves for three RR Lyrae star examples with low or medium brightness values from Campaigns 16, 17 and 18 and for 3 eclipsing binaries in Campaign 13, 17 and 18 in Figs.~\ref{fig:exampleLCs} and \ref{fig:exampleLCs2}, respectively. For comparison we also plot the raw and corrected light curves provided by other pipelines. Each column corresponds to one star, whereas each row shows one pipeline, from top to bottom: SAP \citep{sap}, PDCSAP \citep{PDCSAP}, K2SFF \citep[K2 Self-flatfielding][]{k2sff}, EVEREST \citep[EPIC Variability Extraction and Removal for Exoplanet Science Targets,][]{everest} and \texttt{autoEAP}. The gray points depict the raw light curves, if available. In case of our results, beside the raw data, we show the final versions after K2SC correction and detrending. Each of these pipelines were described in more detail in \citet{eap}. Here we do not intend to repeat these introductions, but we note that not all pipelines have photometry available for the later Campaigns.

For brighter stars, all the pipelines work quite well producing good quality light curves similar to \texttt{autoEAP}. Our experience is that the best photometry, which is closest to ours, is the raw photometry provided by EVEREST.

The first example is EPIC 211888680, a faint (\textit{Kp} = 18.828 mag) double-mode RR Lyrae star, which was observed in Campaign 16. Its SAP light curve is dominated by systematics, which were mostly corrected by PDCSAP, but some flux is still missing in the second half of the measurements. The raw K2SFF data shows a continuous fading, which was kept by their polynomial fit-based corrector, while the quality degraded. EVEREST produced a very similar raw light curve, but their Gaussian process-based corrector overfitted the data and removed the intrinsic variations of the star. Our corrected photometry resulted in a light curve with continuous mean brightness and amplitude. The observed scatter is mainly from Poisson noise.

EPIC 212426424 is a brighter (\textit{Kp} = 16.860 mag) fundamental-mode RR Lyrae star in Campaign 17. The star's pulsational properties are totally unrecognizable in its SAP light curve, because the assigned aperture was too small and a significant amount of the incident flux was excluded. PDCSAP improves the light curve quality, but it is not able to fully compensate the undersized aperture. Furthermore, PDCSAP seems to assume that slow variations are caused by blending and not by changes in the fraction of captured flux, and proceeds with subtracting the trends instead of scaling the light curve with it. This introduces strong, but---as the other pipelines reveal---spurious amplitude variation into the light curve. K2SFF provides good quality raw data, but their corrector destroys the light curve shape afterwards. EVEREST was able to find the optimal aperture size, but the Gaussian process recognized the pulsation as systemtics and totally removed it. The corrected \texttt{autoEAP} light curve shows a beating pattern between the pulsation period and the sampling period, which can be used as a measure of quality. As it can be seen, beside a slight amplitude change, which may originate from a longer time scale Blazhko effect, the quality is excellent for this moderately faint target.

The last star is EPIC 211573254, which is a first-overtone pulsator of similar brightness (\textit{Kp} = 16.471 mag) from Campaign 18. Among the three examples, the SAP gives the best light curve here. The observable systematics were almost fully corrected by PDCSAP, but due to the slightly suboptimal aperture, some precision is lost at the end the observations. The raw K2SFF and EVERST light curves are very similar, they even show the same trend. The other common thing is that both systematics correction methods significantly degraded the quality and identified the instrinsic pulsation as a variation caused by the telescope pointing issues.
 
To compare the photometric qualities given by the different pipelines to other kinds of variables stars, we show three other examples in Fig.~\ref{fig:exampleLCs2}, for the eclipsing binaries
EPIC 251456990 (\textit{Kp} = 18.370 mag), EPIC 212567829 (\textit{Kp} = 18.076 mag) and EPIC 211588342 (\textit{Kp} = 16.098 mag), respectively. The results are very similar to that of the RR Lyrae stars, with the best quality light curves being provided by PDCSAP and raw EVEREST. The corrected \texttt{autoEAP} data are as good as the best available data sets from the other pipelines or provide the best solution among all.

We note that some RR Lyrae candidates included in the K2 observing proposals turned out to be non-pulsating stars, with some light curves showing only instrumental variations with little to no periodic signals. Moreover, the quality of some raw light curves vary significantly throughout a given campaign due to various instrumental problems, such as over-corrected background. Finally, in very few cases the K2SC correction failed to produce an improved light curve and degraded the quality compared to the raw data set.
    
\section{Conclusion}
\label{sec:concl}

Originally, our group developed a method to overcome problems caused by strong systematics present in the light curves of K2 RR Lyrae variable stars \citep{eap}, which have periodicities similar to the telescope's roll motion period, and show large-amplitude and sharp light curve shapes. These light curves are not dissimilar from the flux variations caused by changes in the telescope pointing, characterized by slow rolls and fast attitude resets. In this paper we present the improved and automatized version of the extended aperture selection process, in order to speed up the photometry of stars observed in the K2 campaigns.

Although our original goal was to provide photometry for specific targets, we found that our method is able to assign apertures to any kinds of variable stars with high amplitude variations, such as Cepheids, rotational variables or eclipsing binaries.

The automatized EAP (\texttt{autoEAP}) assigns apertures to stars via the following steps: it detects sources photometrically on each image individually; then through an iterative process performs photometry for each target; identifies high-amplitude variable stars and selects pixels that correspond to that source. The raw light curves are then corrected for systematics using Gaussian processes fitted to the time- and position-dependent components of the data using the K2SC software \citep{k2sc}. The remaining instrumental variations are removed using a low order polynomial fitting technique that we developed, where the polynomial is optimized via phase dispersion minimization.

We processed the light curves of thousands of RR Lyrae targets from Campaigns 0--19. 
We provide the raw EAP, the K2SC corrected and the detrended light curves for further analysis. Though we do not process the light curves for all high-amplitude variable stars in all K2 fields, here we publish and release the \texttt{autoEAP} software to the astronomical community to be able to generate their own photometry using our approach.

Our overall conclusion is that for fainter stars most of the times \texttt{autoEAP} provides the best-quality light curve with high consistency, while for bright stars it yields similar results to other pipelines developed for K2 variable star photometry.

\section*{Software and Data Availability}

The \texttt{autoEAP} python package is available on GitHub\footnote{\url{https://github.com/konkolyseismolab/autoeap}} under an MIT License, can be installed via PyPI and version 0.3 is archived in Zenodo \citep{autoeapzenodo}. Given a local target pixel file or just the name of the target, users can create their own \texttt{autoEAP} light curves. The usage of this software is explained in more detail on Github. Moreover, a basic example is provided in the Appendix.

The processed light curves of 3917 RR Lyrae candidate stars can be downloaded from our webpage\footnote{\url{https://konkoly.hu/KIK/data\_en.html}}. These stars were observed in the following Guest Observer Programs: 
GO0055,
GO1018,
GO2040,
GO3040,
GO4069,
GO5069,
GO6082,
GO7082,
GO8037,
GO9916,
GO10037,
GO11111,
GO12111,
GO13111,
GO14058,
GO15058,
GO16058,
GO17033,
GO18033,
GO19033. The program proposals and target lists are available from the repository of K2 approved targets \& programs\footnote{\url{https://keplergo.github.io/KeplerScienceWebsite/k2-approved-programs.html}}.

We processed all proposed targets from each Campaign except for Campaigns 9 and 11.  Both fields were pointed towards the Bulge, and were also split into two halves, effectively doubling the required processing. C9 was split to double the amount of data that could be stored, whereas C11 was stopped and restarted after an initial pointing error.

As these observations saw the densest star fields, the photometry of stars observed in these Campaigns requires advanced solutions, because of the combined effect of high stellar density, large CCD pixels and pointing variations of the spacecraft. Such efforts were made by multiple groups \citep{C9photometry1,C9photometry2}. Nonetheless, \texttt{autoEAP} still works well for stars in less dense parts of the fields or for brighter stars (V$\lesssim$ 15-16 mag) with a small number of relatively close neighbours (up to 5-10 nearby targets within 2.5-pixel radius depending on the brightness). Therefore we provide photometric results for a limited sample from Campaigns 9 \& 11.

\software{
Python \citep{Python3}, 
Numpy \citep{numpy}, 
Scikit-learn \citep{scikit}, 
Lightkurve \citep{lightkurve}, 
Astropy \citep{astropy:2013,astropy:2018}, 
Photutils \citep{photutils}
}

\facility{\textit{Kepler} space telescope \citep{k2mission}, \textit{Gaia} space telescope \citep{Gaia}}

\acknowledgements

Funding for the \textit{Kepler} and K2 missions are provided by the NASA Science Mission directorate. This research was supported by the KKP-137523 `SeismoLab' \'Elvonal grant of the Hungarian Research, Development and Innovation Office (NKFIH), the Lendület LP2014-17 and LP2018-7/2021 grants of the Hungarian Academy of Sciences, and the MW-Gaia COST Action (CA18104). This research made use of Lightkurve, a Python package for \textit{Kepler} and TESS data analysis (Lightkurve Collaboration, 2018). This work has made use of data from the European Space Agency (ESA) mission {\it Gaia} (\url{https://www.cosmos.esa.int/gaia}), processed by the {\it Gaia} Data Processing and Analysis Consortium (DPAC, \url{https://www.cosmos.esa.int/web/gaia/dpac/consortium}). Funding for the DPAC has been provided by national institutions, in particular the institutions participating in the {\it Gaia} Multilateral Agreement. This research has made use of NASA’s Astrophysics Data System.

\appendix
\section{Example Code}

To install the package from prompt, use:
\begin{lstlisting}[language=bash]
pip install autoeap
\end{lstlisting}

To create raw \texttt{autoEAP} photometry for your own targets, a target pixel file is needed. It is easily obtainable from MAST\footnote{\url{https://archive.stsci.edu/missions-and-data/k2}}. Then the command to create raw photometry using python is:
\begin{lstlisting}[language=Python]
import autoeap
yourtpf = '/path/to/your/tpf/ktwoXXX-cXX_lpd-targ.fits'
time, flux, fluxerr = autoeap.createlightcurve(yourtpf)
\end{lstlisting}
Optionally, providing an EPIC number will trigger \texttt{autoEAP} to search and retrieve the appropriate TPF automatically:
\begin{lstlisting}[language=Python]
import autoeap
starname = 'EPIC211532246'
time, flux, fluxerr = autoeap.createlightcurve(starname)
\end{lstlisting}
With this latter command, one can create \texttt{autoEAP} photometry for any K2 variable star.

To apply K2SC correction, the K2SC core package is needed, which can be installed via:
\begin{lstlisting}
pip install george k2sc
\end{lstlisting}
And then, for \texttt{autoEAP} K2SC corrected light curves, run:
\begin{lstlisting}[language=Python]
time, flux, fluxerr = autoeap.createlightcurve(yourtpf,apply_K2SC=True)
\end{lstlisting}

The best results are usually achieved if the final light curve is detrended. The PDM-based polynomial fitting and removal can be activated via:
\begin{lstlisting}[language=Python]
time, flux, fluxerr = autoeap.createlightcurve(yourtpf,apply_K2SC=True,remove_spline=True)
\end{lstlisting}

\bibliography{autoeap.bib}{}
\bibliographystyle{aasjournal}

\end{document}